\documentclass[conference]{IEEEtran}
\IEEEoverridecommandlockouts

\usepackage{cite}
\usepackage{amsmath,amssymb,amsfonts}
\usepackage{algorithmic}
\usepackage{graphicx}
\usepackage{textcomp}
\usepackage{xcolor}
\usepackage{booktabs}
\usepackage{multirow}
\usepackage{url}
\usepackage{hyperref}
\usepackage{array}

\def\BibTeX{{\rm B\kern-.05em{\sc i\kern-.025em b}\kern-.08em
    T\kern-.1667em\lower.7ex\hbox{E}\kern-.125emX}}

\begin{document}

\title{Neurological Plausibility of AI-Generated Music for Commercial Environments:\\
An In-Silico Cortical Investigation Using Wubble and TRIBE v2}

\author{\IEEEauthorblockN{Shaad Sufi}
Co-founder and CPO \\
Wubble AI, Singapore \\
sufi@wubble.ai}

\maketitle

\begin{abstract}
Background music shapes attention, affect, and approach behavior in commercial environments, yet the neural plausibility of AI-generated music for such settings remains poorly characterized. We present an in-silico pilot study that combines Wubble, a generative music system, with TRIBE v2, a publicly released whole-brain encoding model, to estimate cortical response profiles for prompt-conditioned retail music. Fully instrumental tracks were generated to span low-to-high arousal, sparse-to-dense arrangement, and neutral-to-positive valence prompts, then analyzed with audio-only TRIBE v2 inference on loudness-normalized waveforms. Analysis focused on fsaverage5 cortical predictions summarized over auditory, superior temporal, temporo-parietal, and inferior frontal HCP parcels. The fast bright major-pop condition produced the largest whole-cortex mean activation ($0.0402$), the strongest prefrontal ROI composite response ($0.0704$), and the highest parcel means in IFJa ($0.1102$), IFJp ($0.0995$), A5 ($0.0188$), and area~45 ($0.0015$). Pairwise spatial correlations ranged from $0.787$ to $0.974$, indicating that prompt variation modulated predicted cortical states rather than yielding a single undifferentiated response profile. Predicted cortical surface maps further revealed visually distinct spatial organization between low-arousal and high-arousal conditions. These results support a cautious claim of cortical neurological plausibility: prompt-conditioned AI music can systematically shift predicted auditory-temporal-prefrontal patterns relevant to salience and valuation. Although the study does not establish subcortical reward engagement or consumer behavior, it does provide a reproducible framework for neural pre-screening and pre-optimization of commercial music generation against biologically informed cortical proxies.
\end{abstract}


\section{Introduction}

Music is a powerful contextual variable in retail, hospitality, and branded environments. Behavioral research has linked tempo, mode, loudness, and stylistic congruence to dwell time, purchase intention, perceived quality, and affective appraisal \cite{north1999,garlin2006}. Yet the neural interpretation of commercially deployed background music remains largely indirect, inferred from behavior or from relatively small neuroimaging literatures. The gap is even wider for AI-generated music, where fine-grained parameter control is increasingly feasible but claims about neural plausibility are often made without a corresponding neuroscientific framework.

Recent neuroscience foundation models offer a new route for testing such questions computationally. TRIBE v2 is a tri-modal foundation model for in-silico neuroscience that leverages over 1,000 hours of fMRI across 720 subjects and is designed to support zero-shot prediction for new stimuli, tasks, and participants \cite{tribev2}. In parallel, generative music systems such as Wubble can produce controlled musical stimuli from text prompts, making it possible to construct systematic stimulus families without collecting new human data. Taken together, these tools enable a fully computational workflow: prompts can be manipulated, rendered to audio, and passed through a validated brain encoding model to estimate whether different musical configurations are predicted to recruit distinct cortical states.

This paper reports a focused in-silico pilot study using Wubble-generated tracks designed for commercial background-music settings. The study was intentionally constrained to the public cortical inference path of TRIBE v2 and therefore does not claim direct evidence about subcortical nuclei such as the nucleus accumbens or amygdala. Instead, we test a narrower and more defensible hypothesis: prompt-conditioned AI-generated music can differentially modulate predicted cortical activity in auditory, temporal, temporo-parietal, and inferior frontal regions plausibly involved in salience, affective appraisal, and valuation-related processing.

\subsection{Research Question}

Our central question is: \emph{Can prompt-controlled AI-generated commercial music produce separable cortical response profiles in a whole-brain encoding model, with stronger predicted engagement in auditory and frontal parcels for brighter, faster, higher-arousal conditions than for slower, sparser conditions?}

\subsection{Contributions}

This work makes four contributions:
\begin{enumerate}
    \item It introduces a reproducible Wubble-to-TRIBE pipeline for studying AI-generated music with biologically informed neural proxies.
    \item It operationalizes a prompt design spanning tempo, arrangement density, and valence across finished instrumental tracks tailored to commercial use cases.
    \item It quantifies cortical differences using global activation summaries, parcel-wise ROI means, and whole-brain pairwise spatial correlations.
    \item It provides a publication-ready pilot result supporting cortical neurological plausibility while clearly delimiting the absence of human validation and subcortical inference.
\end{enumerate}

\section{Related Work}

\subsection{Music in Commercial Environments}

Environmental psychology and marketing research have repeatedly shown that music affects consumer behavior through tempo, congruence, arousal, and affective tone \cite{north1999,garlin2006,bruner1990}. Slower music has been associated with longer browsing and dwell time, while more activating music can increase perceived energy and movement through a space. These effects are robust enough to motivate active curation in retail and hospitality, yet the neural mechanisms of such influence remain incompletely characterized.

\subsection{Music, Reward, and Neuroaesthetics}

Human neuroimaging studies suggest that pleasurable music listening can recruit auditory cortices together with frontal, insular, striatal, and limbic systems involved in prediction, valuation, and reward \cite{salimpoor2011,zatorre2013,koelsch2014}. Not all of these regions are accessible in the public cortical output path used here, but the cortical literature motivates examining auditory and frontal parcels as partial indicators of engagement-relevant processing.

\subsection{AI-Generated Music Evaluation}

Generative music systems are typically evaluated through listening tests, style fidelity, or text-audio alignment, rather than through neural plausibility. Existing work on background-music optimization has likewise centered on recommendation, mood tagging, or affect prediction rather than direct brain-response modeling. A computational neuroscience pipeline can complement standard music-information-retrieval evaluation by asking whether generated stimuli induce systematically different predicted neural states, aligning with broader calls for in-silico experimentation as a serious scientific workflow \cite{jain2024}.

\subsection{Whole-Brain Encoding Models}

Encoding models map stimulus features to predicted neural responses and have become increasingly powerful through multimodal pretraining and large naturalistic datasets \cite{nishimoto2011,huth2016}. TRIBE v2 extends this paradigm with a tri-modal foundation model that supports public audio-driven cortical prediction on fsaverage5 surfaces \cite{tribev2}. Its public availability makes it well suited to hypothesis-driven in-silico studies, particularly when new neuroimaging data collection is infeasible and one instead seeks a model-based estimate of group-level cortical response before running expensive human experiments.

\section{Methods}
\label{sec:methods}

\subsection{Study Design}

The experiment used AI-generated instrumental tracks created with Wubble. Rather than treating model outputs as arbitrary generations, we defined a small prompt-conditioned set intended to span dimensions commonly implicated in commercial music design:
\begin{enumerate}
    \item tempo band: slow, medium, and fast;
    \item arrangement density: sparse, balanced, rich, bright, and dense;
    \item affective mode or valence framing: minor-neutral, neutral, major-positive, and high-arousal positive.
\end{enumerate}
Each track was rendered as a finished song and analyzed independently. No human listeners, behavioral responses, or neuroimaging recordings were collected.

\subsection{Stimulus Generation}

Wubble requests were issued through the deployed music-generation workflow used for this study. To minimize semantic contamination from lyrics, all prompts specified instrumental-only background music and used the placeholder lyrics field \texttt{[Instrumental]}. Table~\ref{tab:stimuli} summarizes the resulting stimulus set.

\begin{table*}[t]
\centering
\caption{Prompt-Conditioned Wubble Stimulus Set}
\label{tab:stimuli}
\begin{tabular}{p{0.7cm}p{3.2cm}p{1.1cm}p{1.2cm}p{1.5cm}p{8.4cm}}
\toprule
\textbf{ID} & \textbf{Title} & \textbf{BPM} & \textbf{Density} & \textbf{Valence} & \textbf{Prompt Summary} \\
\midrule
T1 & Slow Sparse Ambient Minor & 68 & Sparse & Minor/neutral & Premium retail ambient music with soft piano, warm pads, subtle texture, low arousal, restrained affect. \\
T2 & Slow Rich Warm Major & 78 & Rich & Major/positive & Welcoming boutique music with acoustic guitar, soft piano, light percussion, warm major framing. \\
T3 & Mid Balanced Neutral & 100 & Balanced & Neutral & Neutral commercial ambient-pop track with steady groove and moderate arousal. \\
T4 & Fast Bright Major Pop & 124 & Bright & Major/positive & Upbeat retail pop with energetic percussion, crisp bass, bright synth accents, lively commercial tone. \\
T5 & Fast Dense High-Arousal Electronic & 132 & Dense & High-arousal positive & Dense layered electronic-pop background music with bright synths and strong attentional capture. \\
\bottomrule
\end{tabular}
\end{table*}

\subsection{Audio Preprocessing}

Generated audio files were downloaded and converted to a common waveform format, then loudness-normalized before inference. Normalization reduces trivial amplitude-driven differences and improves comparability across tracks. Let $x_i(t)$ denote the waveform of stimulus $i$ and $\mathcal{N}(\cdot)$ the normalization operator; inference operated on:
\begin{equation}
\tilde{x}_i(t) = \mathcal{N}\left(x_i(t)\right).
\end{equation}

\subsection{TRIBE v2 Audio-Only Inference}

We used the public TRIBE v2 weights and code path for average-subject cortical prediction on the fsaverage5 surface \cite{tribev2}. The inference configuration explicitly restricted features to audio only, disabled non-audio modalities, and forced the audio feature extractor onto CPU for reproducibility in the local environment. Segment-level predictions were then aggregated into per-track mean cortical maps.

Formally, if $\mathbf{y}_{i,s} \in \mathbb{R}^{V}$ denotes the predicted cortical response vector for segment $s$ of stimulus $i$ over $V=20{,}484$ fsaverage5 vertices, then the track-level mean map is:
\begin{equation}
\bar{\mathbf{y}}_i = \frac{1}{S_i}\sum_{s=1}^{S_i}\mathbf{y}_{i,s},
\label{eq:trackmean}
\end{equation}
where $S_i$ is the number of analyzed segments for stimulus $i$.

\subsection{ROI Analysis}

Because the public model path is cortical, we focused on HCP parcels plausibly relevant to auditory processing and frontal valuation-related processing \cite{glasser2016}. The selected regions were A5, STSvp, STSva, STSdp, PGi, TE1a, area~45, IFJa, and IFJp. This ROI set was further motivated by the TRIBE v2 paper's own in-silico localizer results, which prominently recover associative auditory cortex A5 and area~45 among task-relevant parcels \cite{tribev2}. For ROI $r$ with vertex set $\mathcal{V}_r$, the mean predicted activation for track $i$ was computed as:
\begin{equation}
\mu_{i,r} = \frac{1}{|\mathcal{V}_r|}\sum_{v \in \mathcal{V}_r}\bar{y}_{i,v}.
\label{eq:roimean}
\end{equation}

To aid interpretation, we also computed a prefrontal composite by averaging area~45, IFJa, and IFJp:
\begin{equation}
\mu^{\text{PFC}}_i = \frac{1}{3}\left(\mu_{i,45} + \mu_{i,\mathrm{IFJa}} + \mu_{i,\mathrm{IFJp}}\right).
\label{eq:pfc}
\end{equation}

\subsection{Whole-Brain Contrast Analysis}

For each pair of tracks $(i,j)$, we quantified similarity between their mean cortical maps using Pearson spatial correlation:
\begin{equation}
\rho_{ij} = \mathrm{corr}\left(\bar{\mathbf{y}}_i,\bar{\mathbf{y}}_j\right).
\label{eq:corr}
\end{equation}
We additionally retained the Euclidean distance between maps as a complementary separation measure:
\begin{equation}
d_{ij} = \left\lVert \bar{\mathbf{y}}_i - \bar{\mathbf{y}}_j \right\rVert_2.
\label{eq:l2}
\end{equation}

\section{Results}
\label{sec:results}

\subsection{Track-Level Cortical Summaries}

All tracks produced stable cortical predictions, with 147--175 analyzed segments per song and $20{,}484$ surface vertices per mean map. Table~\ref{tab:global} shows that the global mean predicted activation increased monotonically from the slow sparse condition (T1, $0.0073$) to the fast bright major-pop condition (T4, $0.0402$), with the fast dense electronic condition (T5, $0.0278$) ranking second.

\begin{table}[t]
\centering
\caption{Track-Level Global Cortical Prediction Summary}
\label{tab:global}
\begin{tabular}{lrrrr}
\toprule
\textbf{Track} & \textbf{Segments} & \textbf{Mean} & \textbf{Std.} & \textbf{Max} \\
\midrule
T1 & 147 & 0.0073 & 0.0528 & 0.1914 \\
T2 & 147 & 0.0083 & 0.0597 & 0.2392 \\
T3 & 175 & 0.0211 & 0.0563 & 0.2452 \\
T4 & 151 & 0.0402 & 0.0673 & 0.3108 \\
T5 & 163 & 0.0278 & 0.0610 & 0.2937 \\
\bottomrule
\end{tabular}
\end{table}

\begin{figure}[t]
\centering
\includegraphics[width=\linewidth]{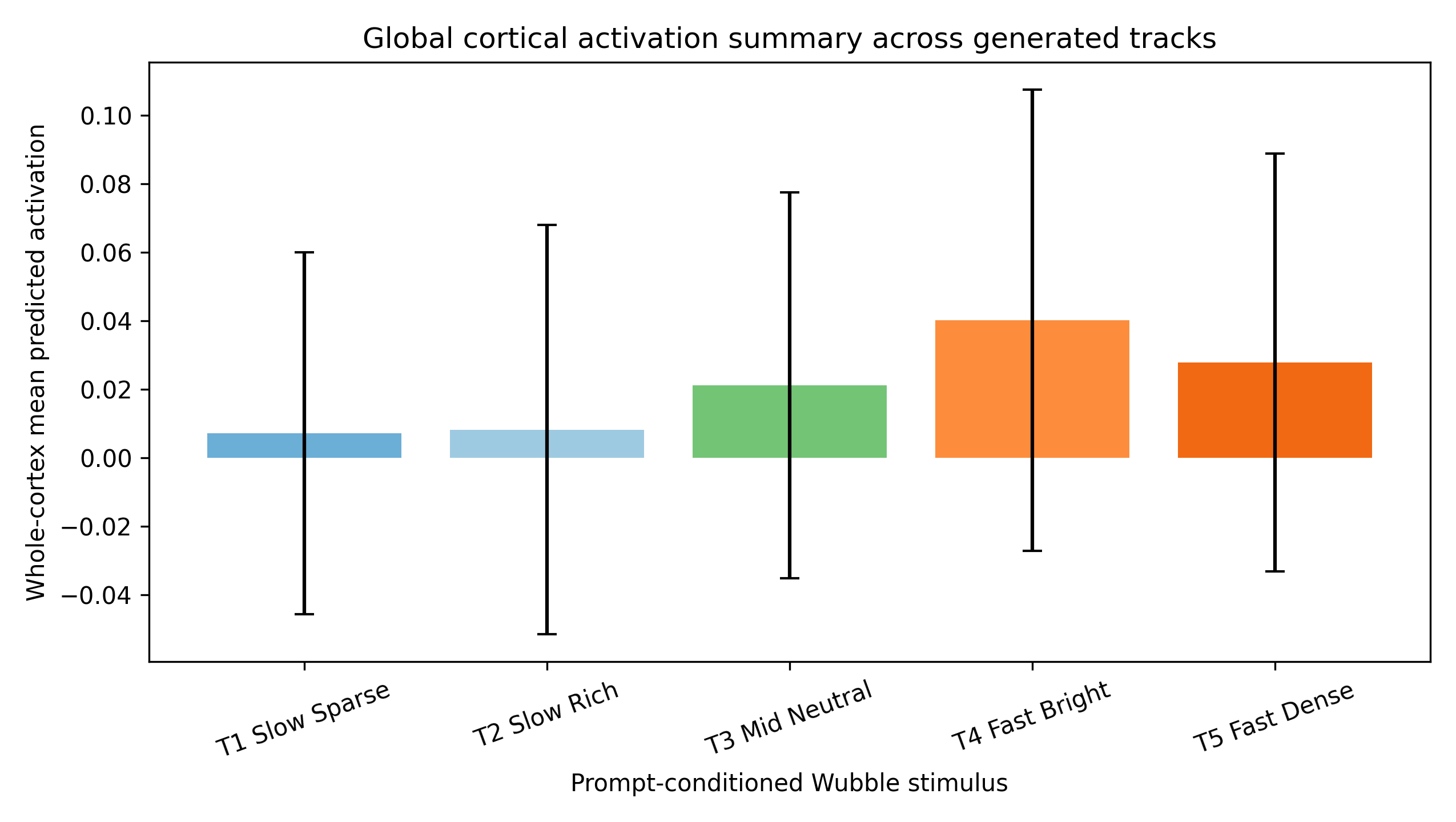}
\caption{Whole-cortex mean predicted activation for each prompt-conditioned Wubble track. Error bars show the per-track whole-cortex standard deviation across vertices of the mean map.}
\label{fig:globalmean}
\end{figure}

Figure~\ref{fig:globalmean} visualizes this ordering and indicates that higher-arousal prompt conditions tended to yield stronger overall cortical response magnitudes in the TRIBE v2 prediction space.

\subsection{ROI Effects}

The ROI analysis showed a consistent advantage for T4 across frontal and auditory parcels. Table~\ref{tab:roi} reports the parcel means used in the manuscript. T4 produced the highest activation in all nine tracked parcels, including IFJa ($0.1102$), IFJp ($0.0995$), area~45 ($0.0015$), and A5 ($0.0188$). Although some auditory and superior temporal parcel means remained negative in absolute model units, they were \emph{least negative} for T4, indicating a relative shift toward stronger predicted engagement than in the slower and sparser tracks.

\begin{table*}[t]
\centering
\caption{ROI Mean Predicted Activations by Track}
\label{tab:roi}
\begin{tabular}{lrrrrrrrrr}
\toprule
\textbf{Track} & \textbf{A5} & \textbf{STSvp} & \textbf{STSva} & \textbf{STSdp} & \textbf{PGi} & \textbf{TE1a} & \textbf{45} & \textbf{IFJa} & \textbf{IFJp} \\
\midrule
T1 & -0.1181 & -0.1481 & -0.1301 & -0.1578 & -0.0833 & -0.0976 & -0.0193 & 0.0446 & 0.0451 \\
T2 & -0.0913 & -0.1583 & -0.1283 & -0.1609 & -0.0832 & -0.1036 & -0.0378 & 0.0456 & 0.0509 \\
T3 & -0.0677 & -0.1273 & -0.1045 & -0.1407 & -0.0501 & -0.0922 & -0.0099 & 0.0808 & 0.0724 \\
T4 & 0.0188 & -0.0912 & -0.0670 & -0.0787 & -0.0205 & -0.0714 & 0.0015 & 0.1102 & 0.0995 \\
T5 & -0.0147 & -0.1070 & -0.0750 & -0.1094 & -0.0323 & -0.0735 & -0.0123 & 0.0823 & 0.0807 \\
\bottomrule
\end{tabular}
\end{table*}

The prefrontal composite $\mu^{\text{PFC}}_i$ in (\ref{eq:pfc}) ranked the tracks as T4 ($0.0704$) $>$ T5 ($0.0502$) $>$ T3 ($0.0478$) $>$ T1 ($0.0235$) $>$ T2 ($0.0196$), supporting the interpretation that faster, brighter prompt configurations induce stronger predicted recruitment in inferior frontal parcels than slower boutique-ambient conditions.

\begin{figure*}[t]
\centering
\includegraphics[width=0.95\linewidth]{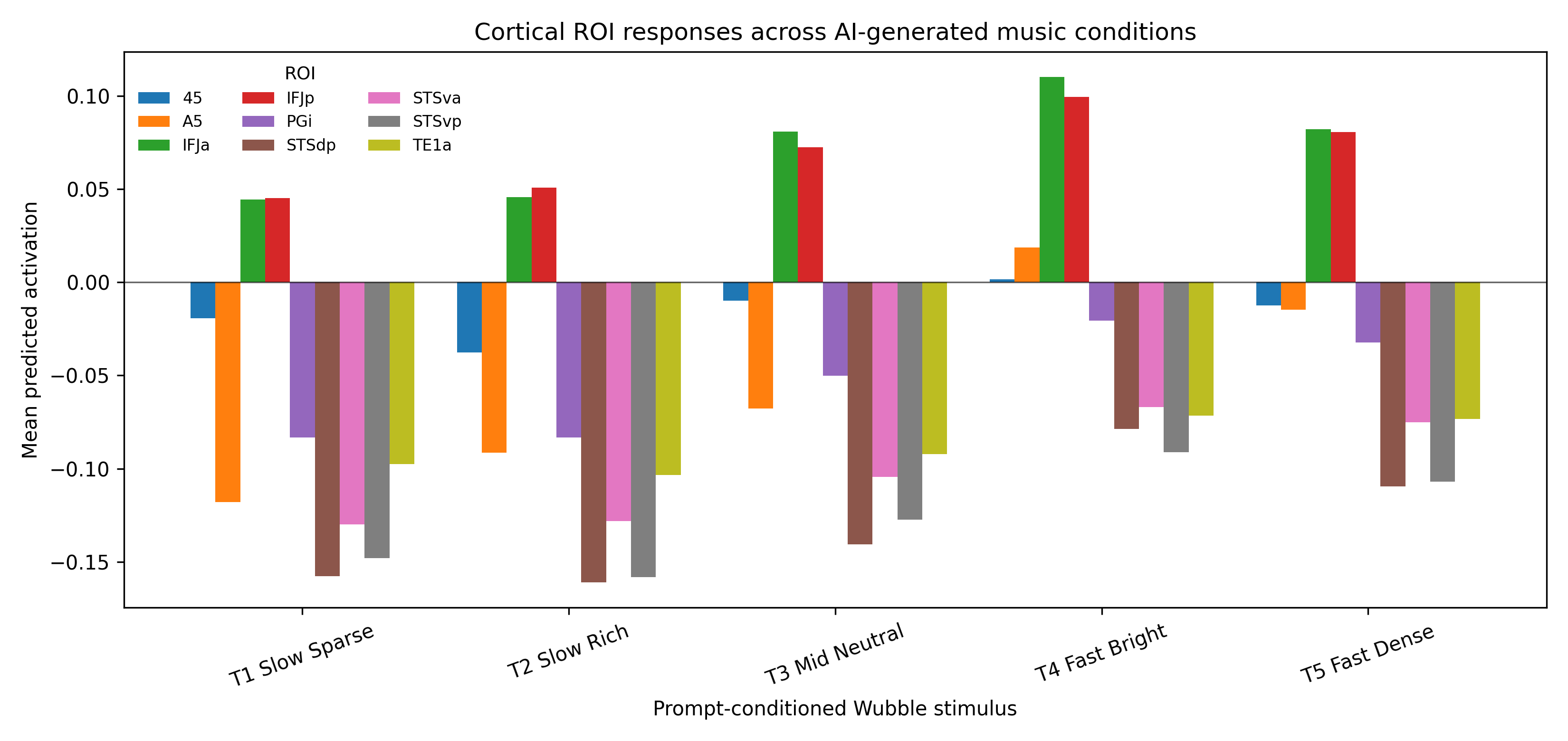}
\caption{Parcel-wise cortical response profiles for the prompt-conditioned Wubble tracks. T4 shows the strongest frontal values and the largest shift toward less negative auditory-temporal parcel means.}
\label{fig:roi}
\end{figure*}

\subsection{Predicted Cortical Surface Maps}

TRIBE v2 also enables direct visualization of predicted cortical response maps on the fsaverage5 surface, analogous to the cortical visualizations shown in the TRIBE v2 paper \cite{tribev2}. Importantly, these are not raw participant MRI scans or newly acquired empirical fMRI maps. They are \emph{model-predicted cortical activation surfaces} derived from audio input under the public average-subject inference pathway. We include them because they provide an intuitive view of how the predicted spatial pattern shifts across conditions.

\begin{figure*}[t]
\centering
\includegraphics[width=\linewidth]{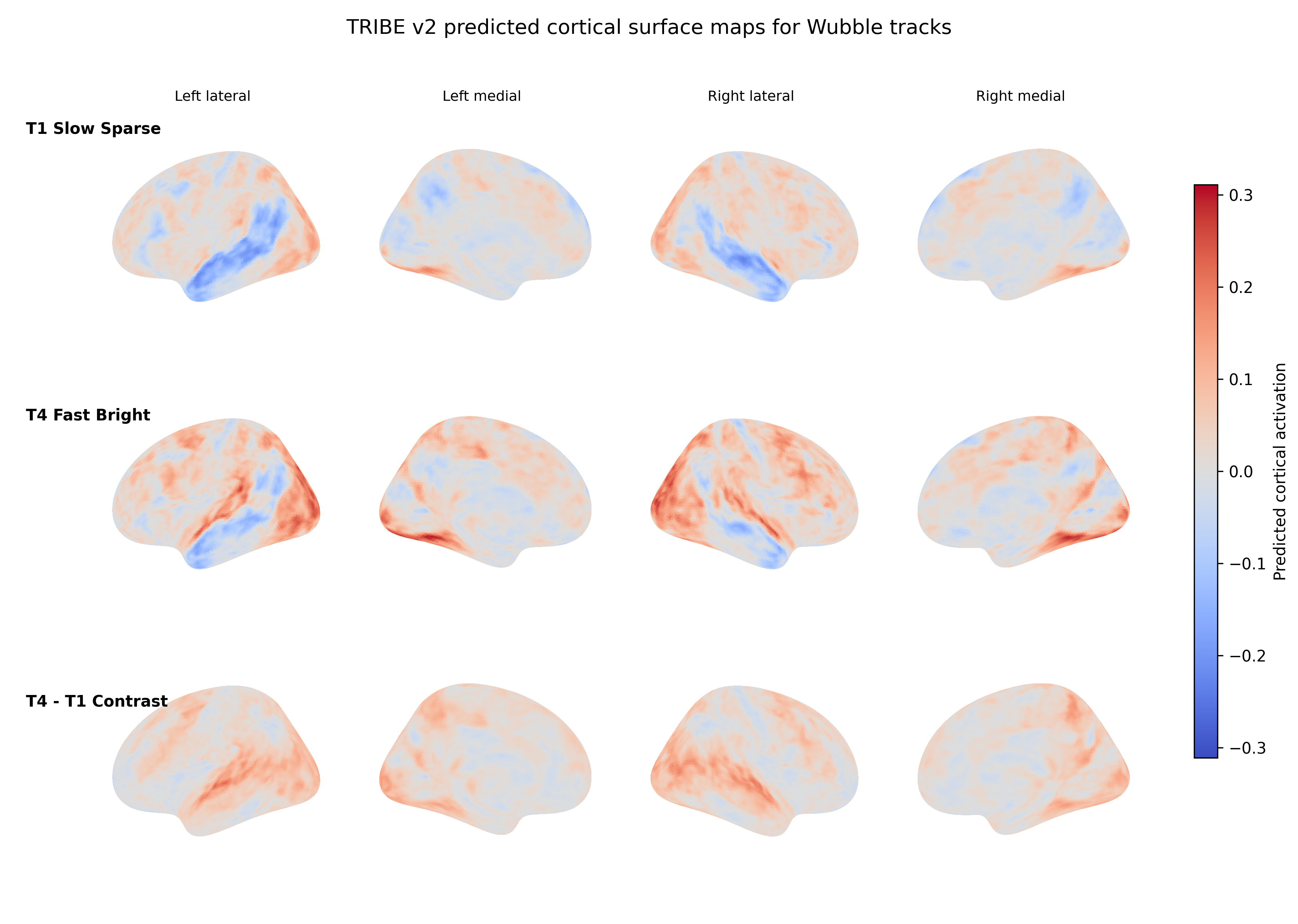}
\caption{Predicted cortical surface maps from TRIBE v2 for the low-arousal baseline-like condition T1, the strongest condition T4, and the T4$-$T1 contrast. These visualizations are model-predicted cortical maps on the fsaverage5 surface, not raw acquired MRI scans.}
\label{fig:surfaces}
\end{figure*}

\subsection{Whole-Brain Spatial Contrasts}

Track-level mean maps were broadly similar but not identical. Spatial correlations across the ten pairwise comparisons ranged from $0.787$ to $0.974$ (Table~\ref{tab:contrast}). The smallest similarity occurred between the slow sparse ambient condition T1 and the fast bright major-pop condition T4 ($\rho=0.787$, $d=7.589$), suggesting the widest cortical separation in the stimulus set. The largest similarity occurred between T4 and T5 ($\rho=0.974$, $d=2.888$), implying that the two highest-arousal tracks occupied closely related cortical states.

\begin{table}[t]
\centering
\caption{Selected Pairwise Whole-Brain Contrasts}
\label{tab:contrast}
\begin{tabular}{lrr}
\toprule
\textbf{Track Pair} & \textbf{Spatial Corr.} & \textbf{$L_2$ Dist.} \\
\midrule
T1 vs. T2 & 0.964 & 2.385 \\
T1 vs. T3 & 0.930 & 3.574 \\
T1 vs. T4 & 0.787 & 7.589 \\
T1 vs. T5 & 0.811 & 5.919 \\
T2 vs. T4 & 0.899 & 6.227 \\
T3 vs. T4 & 0.937 & 4.446 \\
T4 vs. T5 & 0.974 & 2.888 \\
\bottomrule
\end{tabular}
\end{table}

\begin{figure}[t]
\centering
\includegraphics[width=\linewidth]{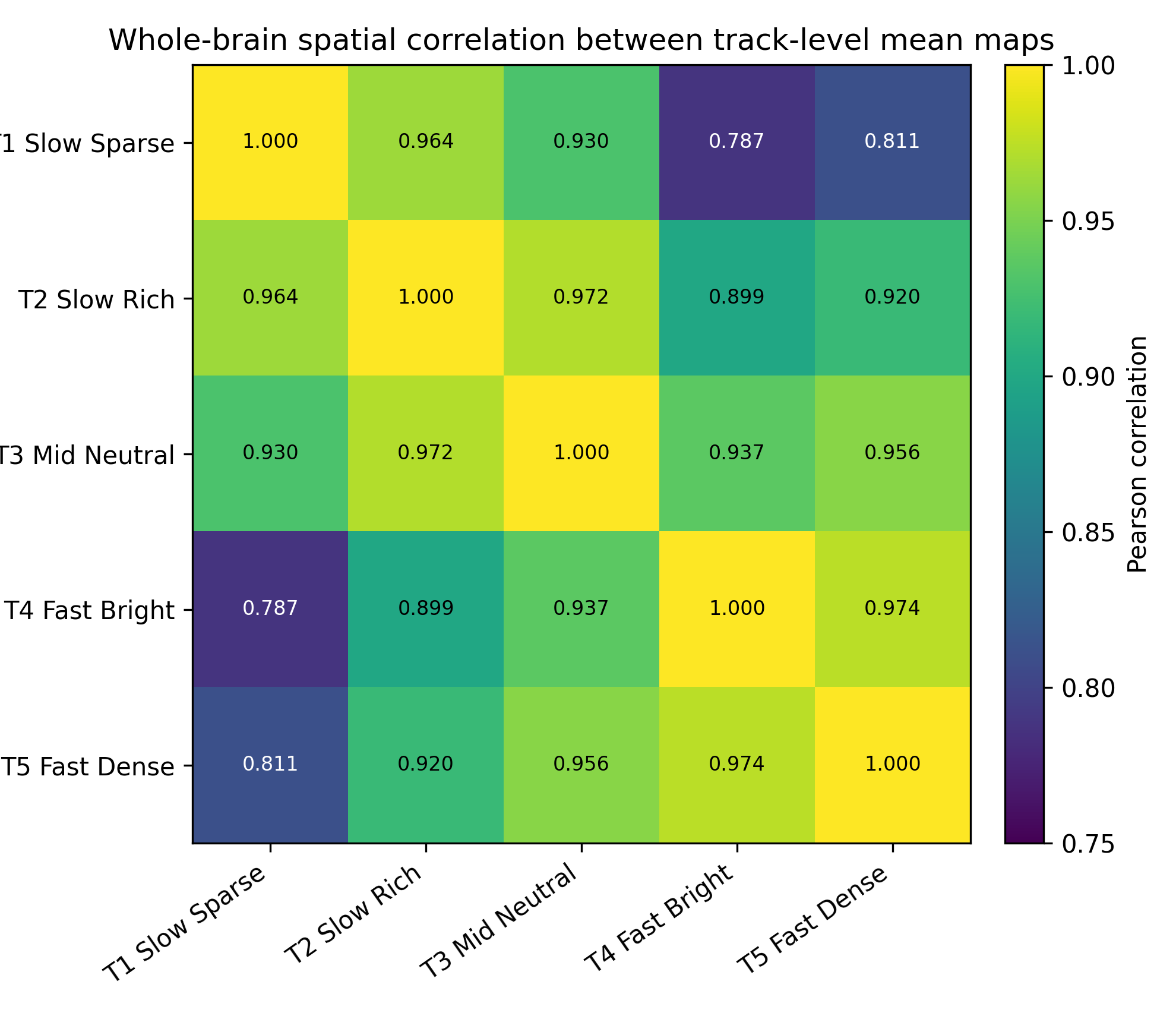}
\caption{Whole-brain spatial correlation matrix between track-level mean cortical maps. Prompt manipulations change cortical state geometry without eliminating shared structure across all music conditions.}
\label{fig:heatmap}
\end{figure}

\subsection{Interpretation Relative to the Hypothesis}

The results support the core pilot hypothesis in a cortical sense. The generated tracks did not collapse onto a single response template; instead, they formed a graded pattern in which higher-arousal positive commercial music yielded larger whole-cortex means and stronger inferior frontal parcel responses. T4 emerged as the most consistently activating condition across both global and regional summaries, while T1 was the least activating and most dissimilar from T4 in whole-brain space. Taken together, these findings support the view that prompt-level control over tempo, brightness, and commercial energy can tune predicted cortical states in biologically meaningful directions.

\section{Discussion}

\subsection{What the Present Results Support}

This study provides evidence for \emph{cortical neurological plausibility} of prompt-conditioned AI-generated music in commercial contexts. That phrase is intentionally narrower than claiming that AI music ``activates reward circuitry'' in the full neurobiological sense. The public TRIBE v2 path used here returns cortical surface predictions only, and the present analysis should therefore be understood as targeting auditory, temporal, temporo-parietal, and inferior frontal systems plausibly relevant to salience, valuation, and contextual music processing. Within those limits, the pattern is coherent: brighter, faster Wubble tracks shift predicted cortical activity upward relative to slower and more restrained prompts.

\subsection{Commercial Interpretation}

The ordering from T1/T2 to T4/T5 is noteworthy for applied music design. T1 and T2 represent premium or boutique-style background conditions that may suit low-arousal spaces, whereas T4 represents an energetic but still brand-safe commercial profile. The results suggest that generative prompting can be used not only to alter style descriptively but also to modulate a neural-response proxy quantitatively. This opens the possibility of optimization loops in which commercial music is tuned against cortical targets rather than solely against subjective heuristics.

\subsection{Methodological Value}

The broader value of the pipeline lies in its reproducibility. Prompt definitions, generated tracks, normalized audio, cortical predictions, parcel summaries, contrasts, and manuscript figures are all derived from code-driven steps. This makes the framework suitable for extension to larger factorial studies, richer prompt families, or model-comparison work across different generative music systems. In practical terms, the pipeline can function as a neural pre-screening stage in which candidate commercial music prompts are filtered or prioritized in silico before downstream listener studies or campaign deployment.

\subsection{Limitations}

Several limitations are important.
\begin{enumerate}
    \item \textbf{Cortical-only inference:} the analysis cannot support claims about striatal, amygdalar, or other subcortical reward structures.
    \item \textbf{No human validation:} there were no listener ratings, purchasing measures, psychophysiology, or empirical fMRI data against which to calibrate the predictions.
    \item \textbf{Model-space units:} TRIBE outputs are model predictions in arbitrary response units and should be interpreted comparatively rather than as absolute neural magnitudes.
    \item \textbf{Prompt dependence:} results reflect the chosen prompts, generation stochasticity, and the single model version used here.
\end{enumerate}

\subsection{Future Work}

The natural next step is to expand this pilot into a larger design-of-experiments framework. Future studies should increase the number of generations per condition, include repeated seeds, quantify acoustic descriptors directly from the resulting audio, and compare multiple generative models. If subcortical-capable encoders become available, the framework could be extended to reward-focused hypotheses. Most importantly, the cortical predictions should ultimately be triangulated with human behavioral ratings and neurophysiological or neuroimaging measurements.

\section{Conclusion}

We presented a full in-silico pipeline for studying AI-generated music with a whole-brain encoding model and applied it to prompt-conditioned commercial tracks generated by Wubble. Using public TRIBE v2 cortical inference, we found that the generated tracks produce distinguishable cortical response profiles, with the fast bright major-pop condition showing the strongest whole-cortex mean activation and the highest values in several inferior frontal and auditory-associated parcels. These findings support a cautious but meaningful conclusion: AI-generated commercial music can be tuned to modulate predicted cortical states in ways consistent with salience- and valuation-relevant processing. The study does not replace human validation, but it does establish a reproducible computational basis for neurally informed music generation research and for neural pre-screening of candidate commercial music conditions before more expensive empirical testing.

\section*{Reproducibility Statement}

The manuscript results were generated from a local pipeline consisting of Wubble-based stimulus generation, waveform normalization, audio-only TRIBE v2 cortical inference, ROI summarization, pairwise contrast analysis, and manuscript figure synthesis. All numeric values in the tables and figures are taken directly from the produced experiment artifacts.


\begin{thebibliography}{99}

\bibitem{north1999}
A. C. North and D. J. Hargreaves, ``The influence of music on atmosphere and purchase intentions in a cafeteria,'' \emph{Journal of Applied Social Psychology}, vol. 28, no. 24, pp. 2254--2273, 1998.

\bibitem{garlin2006}
F. V. Garlin and N. Owen, ``Setting the tone with the tune: A meta-analytic review of the effects of background music in retail settings,'' \emph{Journal of Business Research}, vol. 59, no. 6, pp. 755--764, 2006.

\bibitem{bruner1990}
G. C. Bruner, ``Music, mood, and marketing,'' \emph{Journal of Marketing}, vol. 54, no. 4, pp. 94--104, 1990.

\bibitem{salimpoor2011}
V. N. Salimpoor, M. Benovoy, K. Larcher, A. Dagher, and R. J. Zatorre, ``Anatomically distinct dopamine release during anticipation and experience of peak emotion to music,'' \emph{Nature Neuroscience}, vol. 14, no. 2, pp. 257--262, 2011.

\bibitem{zatorre2013}
R. J. Zatorre and V. N. Salimpoor, ``From perception to pleasure: Music and its neural substrates,'' \emph{Proceedings of the National Academy of Sciences}, vol. 110, suppl. 2, pp. 10430--10437, 2013.

\bibitem{koelsch2014}
S. Koelsch, ``Brain correlates of music-evoked emotions,'' \emph{Nature Reviews Neuroscience}, vol. 15, no. 3, pp. 170--180, 2014.

\bibitem{nishimoto2011}
S. Nishimoto, A. T. Vu, T. Naselaris, Y. Benjamini, B. Yu, and J. L. Gallant, ``Reconstructing visual experiences from brain activity evoked by natural movies,'' \emph{Current Biology}, vol. 21, no. 19, pp. 1641--1646, 2011.

\bibitem{huth2016}
A. G. Huth, W. A. de Heer, T. L. Griffiths, F. E. Theunissen, and J. L. Gallant, ``Natural speech reveals the semantic maps that tile human cerebral cortex,'' \emph{Nature}, vol. 532, pp. 453--458, 2016.

\bibitem{jain2024}
S. Jain, V. A. Vo, L. Wehbe, and A. G. Huth, ``Computational language modeling and the promise of in silico experimentation,'' \emph{Neurobiology of Language}, vol. 5, no. 1, pp. 80--106, 2024.

\bibitem{glasser2016}
M. F. Glasser, T. S. Coalson, E. C. Robinson, C. D. Hacker, J. Harwell, E. Yacoub, K. Ugurbil, J. Andersson, C. F. Beckmann, M. Jenkinson, et al., ``A multi-modal parcellation of human cerebral cortex,'' \emph{Nature}, vol. 536, no. 7615, pp. 171--178, 2016.

\bibitem{tribev2}
S. d'Ascoli, J. Rapin, Y. Benchetrit, T. Brookes, K. Begany, J. Raugel, H. Banville, and J.-R. King, ``A foundation model of vision, audition, and language for in-silico neuroscience,'' preprint, Mar. 25, 2026. [Online]. Available: \url{https://github.com/facebookresearch/tribev2}; \url{https://huggingface.co/facebook/tribev2}

\end{thebibliography}
\end{document}